\documentclass[twocolumn,showpacs,amsmath,amssymb,prd,nofootinbib]
{revtex4}
\def\beq{\begin{equation}}
\def\eeqno#1{\label{#1}\end{equation}}
\def\sss{\scriptscriptstyle}
\def\^#1{^{\sss #1}}
\def\_#1{_{\sss #1}}

\def\rar{\rightarrow}

\def\az{a_0}

\def\cmss{{\rm cm~s^{-2}}}
\def\gev{{\rm Gev}}
\def\mev{{\rm Mev}}

\def\mpc{{\rm Mpc}}
\def\kpc{{\rm kpc}}
\def\cm{{\rm cm}}

\def\s{\sigma}
\def\a{\alpha}
\def\b{\beta}

\def\m{\mu}
\def\n{\nu}
\def\c{\gamma}

\def\d{\delta}
\def\c{\gamma}

\def\kpc{{\rm kpc}}
\def\mpc{{\rm Mpc}}

\def\cd#1{{}_{\sss;#1}}
\def\F{\mathcal{F}}
\def\K{\mathcal{K}}
\def\L{\mathcal{L}}
\begin{document}

\title{Gravitational Cherenkov losses in MOND
theories}
\author{Mordehai Milgrom} \affiliation{DPPA, Weizmann Institute of
Science, Rehovot 76100, Israel}

\begin{abstract}
Survival of high-energy cosmic rays (HECRs) against gravitational
Cherenkov losses is shown not to cast strong constraints on MOND
theories that are compatible with general relativity (GR): theories
that coincide with GR in the high-acceleration limit. The
energy-loss rate, $\dot E$, is shown to be many orders smaller than
those derived in the literature for theories with no extra scale.
The gravitational acceleration produced by a HECR in its vicinity is
much higher than the MOND acceleration $\az$. So, modification to
GR, which underlies $\dot E$, enters only beyond the MOND radius of
the particle, within which GR holds sway: $r\_M=(Gp/c\az)^{1/2}$.
The spectral cutoff, which enters $\dot E$ quadratically, is thus
$r\_M^{-1}$, not the particle's, much larger, de Broglie wavenumber:
$k\_{dB}= p/\hbar$. Thus, $\dot E$ is smaller than published rates,
which use $k\_{dB}$, by a factor $(r\_Mk\_{dB})^2\approx
10^{39}(cp/3\times 10^{11}\gev)^3$.
\par
With $r\_M^{-1}$ as cutoff, the distance a HECR can travel without
major losses is $q\ell\_M$, where $\ell\_M=c^2/\az$
is the MOND length, and $q$ is a dimensionless function of
parameters of the problem. Since $\ell\_M$ is $\approx 2\pi$ times
the Hubble distance, survival of HECRs does not strongly constrain
GR-compatible, MOND theories. Such theories also easily satisfy
existing preferred-frame limits, inasmuch as these limits are gotten
in high-acceleration systems. I exemplify the results with MOND
adaptations of Einstein-Aether theories.
\end{abstract}
\maketitle

\section{Introduction}
Several modifications of general relativity (GR) have been
considered in recent years that involve nonluminal gravitational
wave propagation. Depending on the theory, and on the choice of its
parameters, these nonluminal waves can propagate faster or slower
than light. Examples of such theories are Einstein-Aether (EA)
theories (e.g., \cite{jm01}, and references therein). Superluminal
propagation is usually frowned upon, because it connotes closed
timelike curves with all their problematics, but dissenting opinions
have also been heard (e.g.,
\cite{bruneton06,bruneton07,babichev08,geroch10}). Subluminality, on
the other hand, may be in conflict with the observation of
high-energy cosmic rays (HECRs) that are thought to arrive to earth
from astrophysical distances. Caves \cite{caves80}, and, more
recently, Moore and Nelson \cite{mn01}, have pointed out that unless
the subluminal speed is very near the speed of light, such HECRs,
which themselves move very near the speed of light, lose their
energy very efficiently by gravitational Cherenkov radiation (GCR)
of the subluminal waves. Elliott et al. \cite{ems05} then discussed
the constraints that such considerations cast on the parameters of
EA theories.
\par
The relativistic formulations of MOND proposed to date, in the form
of modified-gravity, involve gravitational degrees of freedom in
addition to the metric.\footnote{This is not necessarily the case in
"modified inertia" versions.} These theories include TeVeS and its
predecessors (\cite{bek04,sagi10}, reviewed in \cite{skordis09}),
adaptations of EA theories to MOND \cite{zlosnik07}, and BIMOND
\cite{milgrom09}. Wave propagation in BIMOND has not been studied
yet, but in the other MOND theories there are nonluminal modes. So,
either parameters can be chosen so that these waves are always
superluminal, or else, these parameter need to be subjected to the
constraints from HECR survival.
\par
I point out here that results as in refs. \cite{mn01,ems05} cannot
be applied to constrain MOND theories, as has been stated before,
without major corrections that greatly weaken the resulting
constraints.
\par
In section II, I show the need for a new derivation of the loss rate
in MOND theories, and obtain the new estimate.
In section III, I exemplify the results with MOND adaptations of EA
theories.

\section{MOND adjusted loss rates}
Detailed calculations of the GCR energy-loss rates by HECRs have
been described in \cite{caves80,mn01,ems05}, for modifications of GR
with no extra scale. Cutting through the details, we may note that
as the electromagnetic Cherenkov emission is proportional to the
square of the particle charge, so the GCR rate is proportional to
$G(\c M)^2=G(p/c)^2$, where $G$ is Newton's constant, and $p$ and
$\c$ are the particle's momentum and Lorentz factor. Then, on
dimensional grounds, the energy radiated per unit time per unit
wavevector, $k$, is
 \beq \frac{d^2E}{dk dt}= \frac{SGp^2k}{c}, \eeqno{spec}
where $S$ is some dimensionless function of parameters of the
theory, such as the wave speed for the particular mode (in units of
$c$), of $\hbar k/p$, and of the particle's $\b=v/c$. To get the
total energy-loss rate, this expression is integrated over $k$, up
to some cutoff. In \cite{mn01,ems05} this is $k\_{dB}= p/\hbar$. The
resulting expression for the loss distance is
 \beq D^{-1}_{loss}\equiv E^{-1}\frac{dE}{dx}=
 \frac{Gp^3}{\hbar^2 c^3}Q,
   \eeqno{rate}
where $Q$ is a dimensionless quantity that depends on dimensionless
parameters of the theory, the wave mode under consideration and on
the speeds of the particle and the wave in units of $c$. The
coefficient $Q$ depends also on the internal structure of the
particle (e.g., partonic structure of a hadron). This enters because
the radiation process assumes coherence over the radiating particle,
but for a high-energy hadron, the length $k\_{dB}^{-1}$ already
probes its internal structure; so coherence applies only to the
smaller constituents. In \cite{mn01} this is taken into account by
considering radiation from partons that each carry 0.1 of the hadron
momentum. This increases $D_{loss}$ by a factor $10^3$ over that for
the whole hadron; in \cite{ems05} the structure is treated in more
detail. Equipped with expression (\ref{rate}), one uses lower limits
on the travel distance, $D$, of high $p$ cosmic rays, to impose
$D_{loss}\gtrsim D$, thus placing upper limits on $Q$, and hence on
the parameters of the theory.
\par
Here I note that the situation in relativistic MOND modifications of
GR is very different, because they introduce a new dimensioned
constant: the MOND acceleration $\az\approx 1.2\times 10^{-8}\cmss$.
I consider only MOND theories that coincide with GR in the limit
$\az\rar 0$--called here GR-compatible, MOND theories. This can be
straightforwardly implemented in the MOND adaptations of EA
theories, in BIMOND, and in the theories discussed in
\cite{halle08}. It is not quite the case for TeVeS, where a
departure from GR remains even in the limit (see, e.g.,
\cite{bek04,sagi09}). In GR-compatible theories we expect an
attenuation of the GCR in circumstances where the gravitational
accelerations are larger than $\az$. For example, a particle
traversing the inner solar system, where the gravitational
acceleration due to the sun is much higher than $\az$, would emit
strongly attenuated GCR, except, perhaps, at wavelengths longer than
the MOND radius of the sun,
$r\_{M,\odot}=(M_{\odot}G/\az)^{1/2}\approx 10^{17}\cm$.
\par
The shorter-wavelength ``anomalous'' waves, whether subluminal or
superluminal, have a much suppressed interaction with matter in
high-acceleration backgrounds, such as the inner solar system, or
the vicinity of close binary pulsars. So, their detectibility, and
emissivity, in such systems is greatly suppressed.
\par
The high-acceleration regime is also very pertinent for hadrons: The
gravitational acceleration produced by a hadron of mass $M$, at
rest, at a distance $\hbar/Mc$, is $\approx
2.6\times10^4(M/M_p)^3\az$, where $M_p$ is the proton mass. For a
high-energy hadron of Lorentz factor $\c$, the near-field,
gravitational acceleration it produces at the distance
$k\_{dB}^{-1}$ is a factor $\c^3$ larger, which is many orders
of magnitude larger than $\az$.\footnote{I am speaking here of the
near field of the hadron, not the radiation field.}
\par
It is impracticable to calculate exactly the GCR loss rate in any
particular MOND theory, let alone give a general result. We can,
however, make an estimate based on the following argumentation: I
surmise that for GR-compatible, MOND theories, there are no
Cherenkov effects within the MOND radius of the particle, where GR
holds sway:
 \beq r\_M=(Gp/c\az)^{1/2}\approx 3\times 10^{-12}
 (cp/\gev)^{1/2}\cm. \eeqno{jupa}
Effectively, the particle carries around it a bubble of radius
$r\_M$, devoid of the subluminal waves. Waves with wavenumbers
larger than $r\_M^{-1}$ are then not produced coherently. I also
assume that in the MOND regime, far beyond $r\_M$, the theory has no
further scale, and so we can use the results from the literature for
the gravitational Cherenkov spectrum for such theories. This means
that we can adopt the Cherenkov spectrum discussed above, but cut it
off above the wavenumber $\sim r\_M^{-1}$, instead of at $k_{dB}$.
The MOND radius is many orders of magnitude larger than
$k\_{dB}^{-1}$: $r\_Mk\_{dB}\approx 160(cp/\gev)^{3/2}$. In the MOND
context the rates appearing in the literature should thus be
reduced, nominally, by a factor $\sim (r\_Mk\_{dB})^2\approx
10^{39}(cp/3\times 10^{11}\gev)^3$, normalized to the momentum used
in the constraints in \cite{mn01,ems05}.
\par
More precisely, to account for MOND effects we may approximately
include in expression (\ref{spec}), a factor $1-\m(a/\az)$, where,
(for $a\gtrsim\az$) $a/\az=(kr\_M)^2$ measures the acceleration at
the radius $k^{-1}$ probed by wavenumber $k$, and $\m$ is a MOND
interpolating function appropriate for the circumstances. So the
factor $1-\m$ measures the departure from GR at a distance $k^{-1}$.
The integral over $k$ now converges if $1-\m(x)$ vanishes at high
$x$, faster than $x^{-1}$, which is known to be required, anyhow,
from solar-system constraints. The MOND factor effectively cuts off
the integral at $k=r\_M^{-1}$. So, integrating the Cherenkov
radiation spectrum, Eq.(\ref{spec}), up to the MOND cutoff, instead
of the de Broglie wavenumber, expression (\ref{rate}) is replaced
by\footnote{Since the $k$-cutoff is now much smaller than $p/\hbar$,
we ignore the dependence of $S$ on $k\hbar/p$, which can be set to 0
in the argument.}
 \beq D_{loss}=q\ell\_M.  \eeqno{ratebh}
Here, $\ell\_M\equiv c^2/\az$ is the MOND length, and $q$, which has
a similar role to that of $Q^{-1}$ in Eq.(\ref{rate}), depends on
parameters of the theory and on the wave and particle velocities (in
units of $c$).\footnote{The factor $q$ insures, for example, that
$D_{loss}$ is infinite if the particle speed is below the wave
speed.} Observations that place a lower limit on $D_{loss}$ then
place lower limits on $q$. As has been noted before in other
contexts, $\ell\_M\approx 2\pi D_H$, where $D_H$ is the Hubble
distance; so the lower limits on $q$ are of order unity, even if we
require survival over $D_H$.
\par
For low-rest-mass particles, such as electrons, photons, and
neutrinos, whose de Broglie wavelength can be larger than their MOND
radius, high-acceleration effects enter only if $ r\_M
k\_{dB}\gtrsim 1$, which is the case for energies
$E>E_M\equiv(c^4\hbar^2\az/G)^{1/3}=E_p(t_p\az/c)^{1/3}\sim 35
\mev$, where $E_p$ and $t_p$ are the Planck energy and time,
respectively. Then our expression above for $D_{loss}$ applies. For
lower energies, $r\_M k\_{dB}<1$, and it is $k\_{dB}$ that sets the
cutoff. Then, the expression for $D_{loss}$ is multiplied by
$(E_M/E)^3$.
\par
Since the MOND radius is much larger than the size of a hadron, the
particle radiates coherently, and there is no need to account for
its internal (parton) structure. So, after correcting the rates for
scaleless theories by a factor $(r\_M k\_{dB})^2$, to make them fit
for MOND theories, we also have to correct back the partonic effects
on the rates. For example, in \cite{mn01}, $Q$ defined in
Eq.(\ref{rate}) is $Q\sim 10^{-3}(n-1)^2$, where $n$ is the ratio of
$c$ to the wave speed (the particle speed is assumed much nearer to
$c$), and the factor $10^{-3}$ assumes that the relevant partons
carry 0.1 of the hadron momentum $p$. In the MOND context we would
have, instead, $q^{-1}\sim (n-1)^2$, where the factor $10^{-3}$ has
been removed. So even for travel distances of the Hubble length, the
bound on $n-1$ is only of order 1. (The expression for $Q$ in
\cite{mn01} assumes $n-1\ll 1$; so a more careful treatment is
needed in general.)
\par
The effect I propose here is analogous to that calculated for the
electromagnetic Cherenkov case, when a charged particle moves along
the axis of a vacuum cylinder of radius $R$ (with refractive index
$n=1$) surrounded by a medium with $n>1$. Since the medium with
$n>1$ is kept a distance $R$ from the particle, the wavenumber
spectrum is, indeed, found to be sharply cutoff above $\sim R^{-1}$
(see e.g., \cite{bol61}). This happens also for a particle moving in
vacuum, parallel to the interface of a half-space dielectric, with
$n>1$, a distance $R$ away. Perhaps a better analogy is when the
particle, which moves through a medium with $n>1$, induces around
it, and carries with it, a region of radius $R$ (in the medium's
rest frame) where $n=1$ to a high accuracy. My result is also akin
to the effect of a finite radius, $R$, of the emitting charge (e.g.,
\cite{zak82,smith93}), where the radiation spectrum is also sharply
cut off for $kR>1$.
\par
There is, clearly, a more general lesson to be learnt from our
analysis. For example, similar corrections would apply to other
GR-compatible modifications involving a new scale, such as
modifications of GR at large distances.
\par
Note, in passing, that constraints from solar-system and
binary-pulsar limits on preferred-frame parameters, are easily
satisfied by GR-compatible, MOND theories (not by TeVeS,
\cite{sagi09}). These limits, inasmuch as they pertain to
experiments in high-acceleration regions, test such theories where
they differ from GR by only $\sim 1-\m(a/\az)$, which can be
extremely small. Note also, that in the MOND limit of such theories
($\az\rar \infty$, with $G\az$ fixed), the post-Newtonian expansion
is not applicable, since the nonrelativistic (NR) limit of the
theory is then not Newtonian, contrary to the basic premise of the
post-Newtonian expansion.

\section{Example: MOND adaptation of Einstein-Aether theories}
Consider, as a specific example, the MOND formulations that were
adapted from EA theories \cite{zlosnik07}. In EA theories
(\cite{jm01}, and references therein), gravity is carried by a
metric, $g_{\m\n}$, as well as a vector field, $A^\a$. One adds to
the standard Einstein-Hilbert Lagrangian for the metric, the terms
(in this section I take $c=1$) \beq \L(A,g)=\frac{\az^2}{16\pi G}\K+
\L_L,\eeqno{lagrangeII}
 \beq
\K=\az^{-2}\K^{\a\b}_{\c\s}A^{\c}\cd{\a}A^{\s}\cd{\b},\eeqno{KKK}
\beq \K^{\a\b}_{\c\s}=c_1g^{\a\b}g_{\c\s}+c_2\d^\a_\c\d^\b_\s
+c_3\d^\a_\s\d^\b_\c+c_4A^\a A^\b g_{\c\s}, \eeqno{mutas}
 and $\L_L$ is a
Lagrange multiplier term that forces the vector to be of unit
length.
\par
Elliott et al. \cite{ems05} considered three
constraints on the coefficients, resulting from GCR losses to
the different modes.
\par
These EA theories have been adapted to give MOND phenomenology in
the NR regime \cite{zlosnik07}, by modifying the first
term in the added Lagrangian to
 \beq \L_M(A,g)=\frac{\az^2}{16\pi G}\F(\K).\eeqno{lagrange}
Zlosnik et al. \cite{zlosnik07} limited themselves to theories with
$c_4=0$. They showed that to get the correct NR MOND phenomenology
we have to have $\F(\K)\approx \K+\eta\K^{3/2}$ for $\K\ll 1$, and
$c_1=-2$ (with their sign convention for the $c_i$, which is
opposite to that of \cite{ems05} and of others).\footnote{This makes
the speed of one of the modes--the trace aether-metric mode--vanish,
assuming no background acceleration field, $\nabla\phi_b$. If our
system is in such a field, e.g., in the vicinity of galactic
systems, $(v/c)^2$ of the different waves may suffer additive
corrections of order $|\nabla\phi_b|/\az$, when
$|\nabla\phi_b|/\az\lesssim 1$.}$^,$\footnote{A constant factor $\a$
in front of $\K$ on the right-hand side, as appears below Eq.(10) of
\cite{zlosnik07}, is absorbed in the $c_i$, and is taken $\a=1$.} In
addition, if we want the NR limit of the theory to tend to Newtonian
gravity for $\az\rar 0$, as it should, we need to have
$\F'(\K\rar\infty)\rar 0$. In fact, $\F'(\K)$ plays the role of
$1-\m$ in the discussion leading to Eq.(\ref{ratebh}). To make the
theory GR compatible, I require, to be on the safe side, that $\F$
is bounded by a constant $\zeta$. Matter (and the metric) then
decouples from the vector field in the limit, and is governed by
GR.\footnote{Thus, vector waves will be very difficult to detect in
the inner solar system, if they have $k^{-1}<r\_{M,\odot}$.} (If
$\F\rar\zeta\not =0$, we get GR with a cosmological constant $\sim
\zeta\az^2$.) Here, as in \cite{zlosnik07}, $G$ is Newton's
constant. If we replace it with $G'\not =G$, the NR limit can still
be made a good MOND theory. To then get the Newtonian limit for
$\az\rar 0$, we have to have $\F'(\K\rar\infty)\rar \d\not=0$, with
an appropriate $\d$. But then, the relativistic theory seems not to
be GR compatible. A similar situation occurs in BIMOND
\cite{milgrom09}.\footnote{Interestingly, MOND adaptations of EA are
constrained versions of the BIMOND class: If we force the two
metrics $g\^{\m\n}$ and $\hat g\^{\m\n}$, which in BIMOND are
independent degrees of freedom, to be related by $\hat
g\^{\m\n}=g\^{\m\n}+sA^\m A^\n$ ($s<1$), with, furthermore,
$g\_{\m\n}A^\m A^\n=-1$, we get the EA-MOND class of theories, with
$g\_{\m\n}$ and $A^\m$ as degrees of freedom. For example, with
these constraints, the BIMOND scalar arguments
$\Upsilon,~\hat\Upsilon$, shown in \cite{milgrom09} to lead to
particularly simple theories, give EA-MOND versions with
$c_1=c_4=-c_3$, while $c_2$ is free and introduced by scalars of the
form $g\_{\m\n}\bar C^\m\bar C^\n$, and its image under metric
interchange, terms which retain the simplicity.}
\par
It is impracticable to get an exact solution of the full problem of
GCR, especially as the problem involves the full acceleration range,
as we saw above. As a very reasonable approximation I proceed as
follows. In the field equations, the $c_i$, when they are outside
the argument of $\F$, appear always multiplied by $\F'(\K)$. So, in
high-acceleration environment, where $\F'(\K)\rar 0$, we are in a
situation that can be reproduced by taking the limit $c_i\rar 0$,
with their ratios fixed, in a standard (scale free) EA theory. In
this limit, all the loss rates calculated in \cite{ems05} vanish, as
expected. This implies that the GCR losses are, indeed, cut off for
waves with $k>k\_M=r\_M^{-1}$. (As explained earlier, the cutoff is
not sharp, but since the Cherenkov spectrum increases only as $k$,
the cutoff is effectively at $k\_M$.)
\par
Next, we note that in the MOND regime--where $\K\ll 1$, and so
$\F(\K)\approx \K$--we get an EA theory, as studied by \cite{ems05}.
\par
I thus approximate the rates in this MOND adaptation of EA theories,
by making two corrections to the expressions for the rates in
\cite{ems05}: First I replace their cutoff wavenumber by
$r\_M^{-1}$, as described in section II. Then I increase the rate by
a factor of $10^3$ to account for the fact that with the smaller
cutoff, the hadron radiates coherently, and not as the sum of its
partonic contributions. This factor corrects back the partonic
correction of \cite{mn01}.\footnote{Elliott et al. \cite{ems05} say
that their more detailed treatment of the structure gives a somewhat
stronger constraint than that of \cite{mn01}; this means that our
correcting back in this way is conservative.}
\par
The constraints derived in \cite{ems05} are in the form of upper
limits on three functions of the $c_i$, $f_k(c_i),~k=1,2,3$,  with
coefficients of order 1 (one for each mode), assuming that HECRs
travel a distance $D$, of at least 10\kpc, without major losses.
These functions vanish when $c_i=0$, and are, generically, of
order 1 when the $c_i$ are of order 1. Only the forms of $f_k$ for
$c_i\ll 1$ are given, because the constraints in \cite{ems05}
require this; but this is not the case for MOND theories. The
constraints in \cite{ems05} for standard EA theories [their
eqs.(3.13) (squared here), (4.7), and (4.15)]
are:\footnote{Here I normalize the constraints of \cite{ems05} to a
travel distance of $D=10\mpc$, instead of their less constraining
$10\kpc$.}
 \beq f_k(c_i)\lesssim b_k(D/10\mpc)^{-1},  \eeqno{theirs}
where $b_1=2.5\times 10^{-34}$, $b_2=7\times 10^{-35}$,
$b_3=10^{-33}.$ With the MOND modification we get instead $b_1=250$,
$b_2= 70$, $b_3=10^3$.  So, even if the coefficients $c_i$ are of
order 1, HECRs can propagate a Hubble distance before they lose much
of their energy, in line with the result in section II.
\par
In addition to their constraints from GCR, \cite{ems05} consider the
emission of a double scalar mediated by a graviton. This process
will not concern us here, as it involves some more delicate
processes and assumptions. In any event, even if this constraint
remains, it is much less difficult to satisfy, as it is only one
constraint on the ratio of the $c_i$, not on their absolute values,
as for the Cherenkov constraints. It is probably also subject to
major relaxation in the MOND context, via the mechanism I discuss
here.

This research was supported, in part, by a center of excellence
grant from the Israel Science Foundation. Discussions with Bob
Sanders and Gilles Esposito-Farese are gratefully acknowledged.

\end{document}